\documentclass{jkas}
\def\year{2016} 
\def\month{June} 
\def\volume{00} 
\def\issue{0} 
\def\beginpage{1} 
\def\endpage{99} 
\def\received{April 1, 2016} 
\def\accepted{April 2, 2016} 
\date{Received \received; accepted \accepted}
\usepackage{dblfloatfix}
\def\simlt{\lower.5ex\hbox{$\; \buildrel < \over \sim \;$}}
\def\simgt{\lower.5ex\hbox{$\; \buildrel > \over \sim \;$}}
\def\kms{km s$^{-1}$}

\def\feonesix{[Fe II]\,1.644\,$\mu$m}
\def\fefive{[Fe II]\,5.340\,$\mu$m}
\def\feoneseven{[Fe II]\,17.94\,$\mu$m}

\def\fetwofive{[Fe II]\,25.99\,$\mu$m}
\def\fethreefive{[Fe II]\,35.35\,$\mu$m}
\def\fetwo{[Fe II]\ }

\def\hbeta{H$\beta$}

\def\ncr{$n_{\rm cr}$}
\def\tex{$T_{\rm ex}$}

\def\mum{$\mu$m}
\title{Infrared [Fe II] Emission Lines from Radiative Atomic Shocks}
\author[1,2]{Bon-Chul Koo}
\author[3]{John, C. Raymond}
\author[1]{Hyung-Jeong Kim}
\affil[1]{Department of Physics and Astronomy, Seoul National University, Gwanak-gu, Seoul 151-747, Korea; \email{koo@astro.snu.ac.kr, hjkim@astro.snu.ac.kr}}
\affil[2]{Visiting Professor, Korea Institute of Advanced Study, Seoul 02455, Korea}
\affil[3]{Harvard-Smithsonian Center for Astrophysics, 60 Garden Street, Cambridge, MA 02138, USA; 
\email{jraymond@cfa.harvard.edu}}
\def\corrauthor{B.-C. Koo}
\def\runningauthor{B.-C., Koo, J. C. Raymond, H.-J. Kim}

\def\runningtitle{Infrared [Fe II] Emission Lines from Radiative Shocks}

\def\keywords{atomic processes: hydrodynamics: infrared: ISM: shock waves}

\def\abstracttext{
[Fe II] emission lines are prominent in the infrared (IR), and they are important diagnostic tools for   
radiative atomic shocks. We investigate the emission characteristics of [Fe II] lines 
using a shock code developed by \cite{raymond1979} with updated atomic parameters.
We first review general characteristics of IR [Fe II] emission lines 
from shocked gas, and 
derive [Fe II] line fluxes as a function of shock speed and ambient density. 
We have compiled the available IR [Fe II] line observations of interstellar shocks
and compare them to the ratios predicted from our model. 
The sample includes both young and old supernova remnants 
in the Galaxy and the Large Magellanic Cloud and several Herbig-Haro objects. 
We find that the observed ratios of IR [Fe II] lines generally 
fall on our grid of shock models, 
but the ratios of some mid-infrared lines, e.g., \fethreefive$/$\fetwofive, \fefive$/$\fetwofive, 
and \fefive$/$\feoneseven, are significantly 
offset from our model grid. We discuss possible explanations 
and conclude that the uncertainty in atomic rates 
might be the major source of uncertainty, 
while uncertainties in the shock modeling 
and the observations certainly exist.
}

\begin{document}
\jkashead 

\section{Introduction}

Shocks are ubiquitous in the interstellar medium (ISM). 
Supernova (SN) blast waves, 
stellar winds, and outflows/jets from young stellar objects 
are among diverse driving sources. 
The shocks associated with these sources generally 
manifest themselves in various metallic ionic lines,
which may be compared to theoretical models of shock emission lines 
to reveal the physical conditions of their environments 
and also to infer the nature of driving sources. 
Most of these sources, however, are located in the Galactic plane where
the extinction can be so large that optical emission lines  
are not observable.

In radiative atomic shocks, in the 
NIR/MIR band, the forbidden lines from Fe$^+$ are usually most prominent.
There are some unique features that make [Fe II] lines strong
\citep{mckee1984, hollenbach1989b, oliva1989}. 
First, the Fe$^+$ ion has many (16) levels 
with low excitation energies, so that 
these levels are easily excited in shocked gas 
and the transitions among them result in many lines in IR, 
particularly in the NIR band.
Second, the ionization potential of an Fe atom is 7.9 eV ($<13.6$ eV), 
so that there is an extended region behind the shock front 
where Fe is ionized to Fe$^+$ by FUV radiation from the shock front 
while H atoms are mostly  neutral. 
This is in contrast to photoionized regions where Fe atoms   
are mostly in higher ionization stages unless the ionizing radiation 
is hard enough to penetrate deep into interstellar cloud, such as in active 
galactic nuclei. 
Third, the Fe abundance can be enhanced due to grain destruction by shocks  
and can approach its cosmic abundance., i.e., Fe/H$=3.47\times 10^{-5}$ 
by number \citep{asplund2009}, while in the general ISM, 99\% of Fe is locked in dust. 
In supernova remnants (SNRs), 
the newly-synthesized Fe from SN can enhance the Fe abundance too.  
Therefore,  [Fe II] emission lines from shocked gas are stronger than 
those from photoionized regions, e.g. 
[Fe II]/Pa$\beta$=1--10 compared to 0.01 \citep{mouri2000,koo2015}.
That makes these lines very useful for the study of interstellar shocks
\citep[e.g.,][]{dinerstein1995,nisini2008}.

Theoretical calculations of IR [Fe II] lines from interstellar shocks 
have been done by several groups,
i.e., \citet{mckee1984, hollenbach1989b, hartigan2004}.
\cite{mckee1984} calculated [Fe II] 1.257 $\mu$m intensities for 
40 and 100~\kms\ shocks propagating into atomic gas of 
ambient hydrogen nuclei density $n_0=10$~cm$^{-3}$  
and also a 100 \kms\ shock into atomic gas of $n_0=100$~cm$^{-3}$.
\cite{hollenbach1989b} calculated the [Fe II] 1.257 and 1.644 $\mu$m 
intensities for dissociative $J$-shocks with speeds of  
30--150~\kms\ incident upon molecular gas
of $n_0=10^3$--$10^4$~cm$^{-3}$.
The above works included grain destruction.
\citet{hartigan2004} calculated the intensities of strong 
NIR [Fe II] lines for relatively slow shocks (30--50~\kms) 
propagating through atomic gas of $n_0=10^3$--$10^5$ cm$^{-3}$  
appropriate for outflows from protostellar objects. 
Recently, \citet{allen2008} presented a detailed grid of shock models
for $v_s=100$--1,000~\kms\ and $n_0=0.01$--1,000 cm$^{-3}$  
using the shock modeling code MAPPINGS III \citep{dopita1995, dopita1996},
but they do not tabulate the Fe line intensities.

In this paper, we use the shock code of \cite{raymond1979} and \cite{cox1985} 
with updated atomic parameters to model [Fe II] emission lines 
from atomic shocks. One of the motivations of this work is the 
availability of revised atomic constants for the [Fe II] lines. 
Fe$^+$ is a complex ion and its atomic constants, e.g., coefficients of 
collision rates and radiative transitions, are still not accurate. 
Another motivation is the accumulation of IR [Fe II] line 
observations of interstellar shocks, including 1 to 2.5 $\mu$m ground-based
spectra and {\it Spitzer} spectra at longer wavelengths.
It is worthwhile to compare the observed 
[Fe II] emission line parameters of these sources and to confirm their similarities and differences. 
The organization of this paper is as follows.     
In \S~2, we summarize the atomic constants and also the 
parameters of strong [Fe II] lines. We briefly discuss the basic applications of 
[Fe II] lines, as well.  
In \S~3, we investigate the shock structure and the physical properties of the 
[Fe II]-line emitting layers. Physical variables, e.g., temperature and density, 
usually vary through the emitting layer, and we discuss the significance of the 
parameters derived by solving rate equations. We also derive [Fe II] line fluxes 
as a function of shock speed (20--200~\kms) and density (10--$10^3$~cm$^{-3}$), 
and compare them to those of previous studies.  
In \S~4, we present shock grids  
for NIR and MIR [Fe II] emission lines and compare them to 
the available observations of interstellar shocks.  
In \S~5, we conclude our paper.

\section{Atomic Constants and Basic Applications of [Fe II] Lines}

\subsection{Atomic Constants}

\begin{figure}
\centering
\includegraphics[width=80mm]{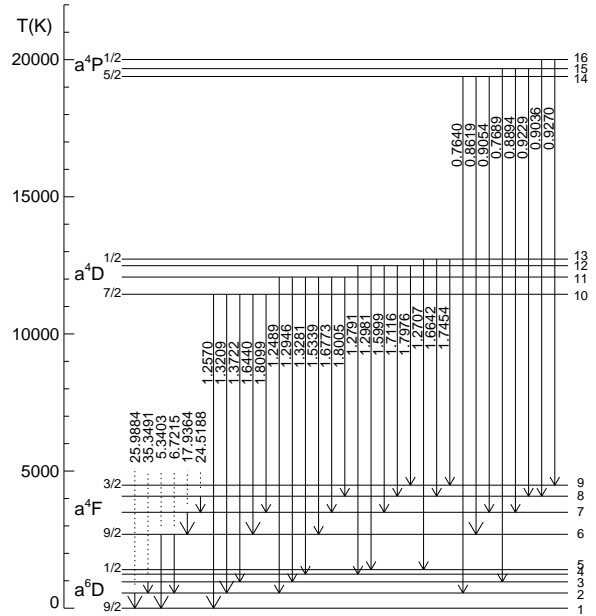}
\caption{
Energy level diagram for the first four low-lying terms of Fe$^{+}$. 
For convenience, we numbered the levels consecutively by 1 to 16 from the ground level. 
The transitions resulting 
strong lines in Table 1 are marked with their wavelengths (in $\mu$m). 
Relatively strong lines 
(i.e., the lines with intensities stronger than 30\% of 
the 1.644 $\mu$m line in Table 1) 
are marked by thick solid lines.
The temperature scale bar on the 
left shows the excitation energies of the levels. 
}
\label{fig1}
\end{figure}

The Fe$^+$ ion has four ground terms, 
a$^6$D (3d$^6$4s), a$^4$F (3d$^7$), a$^4$D (3d$^6$4s), and a$^4$P (3d$^7$), 
each of which has 
3--5 closely-spaced levels to form a 16 level system 
(Pradhan \& Nahar 2011; see Fig. 1). 
The energy gap between the ground level and the excited levels 
is less than $2\times 10^4$~K, so that 
these levels are easily excited in postshock cooling region, and 
the emission lines resulting from the 
transitions among them appear in near- to far-IR bands. 
The energy gap to the next coupled higher-energy term (b$^4$P) is 0.9 eV,
so that it is usually acceptable to consider only these 16 levels 
in computing the line intensities from shocked gas. 
(There are several other terms 
between a$^4P$ and b$^4$P, but they are weakly coupled. See \cite{pradhan2011}.)
There are numerous [Fe II] lines at visual wavelengths, but they are
not included in our models.

\begin{table*}[t]
\caption{Strong [Fe II] lines from transitions among the levels in ground terms}
\centering
\begin{tabular}{rcrrrrrrrrc}
\toprule
\multicolumn{2}{c}{Level ID} & 
\multicolumn{2}{c}{Levels} & 
{\hfil $\lambda$ \hfil} & {\hfil $T_{\rm ex}$ \hfil} & {\hfil $A_{21}$ \hfil} & 
{\hfil $n_{\rm cr}$ \hfil} & 
\multicolumn{3}{c}{Line intensity} \\
u & l & upper & lower & {\hfil ($\mu$m) \hfil} & {\hfil (K) \hfil} & 
{\hfil (s$^{-1}$) \hfil} & {\hfil (cm$^{-3}$) \hfil} &  
$n_e=10^3$ & $10^4$ & $10^5$ cm$^{-3}$ \\
\midrule
 2 &  1 & a6D7/2 & a6D9/2 & 25.9884 &    554 & 2.14e-03 & 3.40e+03 & 0.51 & 0.34 & 0.17 \\
 3 &  2 & a6D5/2 & a6D7/2 & 35.3491 &    961 & 1.58e-03 & 2.28e+03 & 0.11 & 0.10 & 0.06 \\
 6 &  1 & a4F9/2 & a6D9/2 &  5.3403 &  2,694 & 1.30e-04 & 4.44e+02 & 0.99 & 0.30 & 0.07 \\
 6 &  2 & a4F9/2 & a6D7/2 &  6.7215 &  2,694 & 1.16e-05 & 4.44e+02 & 0.07 & 0.02 & 0.00 \\
 7 &  6 & a4F7/2 & a4F9/2 & 17.9364 &  3,496 & 5.86e-03 & 1.53e+04 & 0.24 & 0.44 & 0.41 \\
 8 &  7 & a4F5/2 & a4F7/2 & 24.5188 &  4,083 & 3.93e-03 & 1.05e+04 & 0.05 & 0.11 & 0.13 \\
10 &  1 & a4D7/2 & a6D9/2 &  1.2570 & 11,446 & 5.27e-03 & 3.28e+04 & 1.36 & 1.36 & 1.36 \\
10 &  2 & a4D7/2 & a6D7/2 &  1.3209 & 11,446 & 1.49e-03 & 3.28e+04 & 0.37 & 0.37 & 0.37 \\
10 &  3 & a4D7/2 & a6D5/2 &  1.3722 & 11,446 & 9.72e-04 & 3.28e+04 & 0.23 & 0.23 & 0.23 \\
10 &  6 & a4D7/2 & a4F9/2 &  1.6440 & 11,446 & 5.07e-03 & 3.28e+04 & 1.00 & 1.00 & 1.00 \\
10 &  7 & a4D7/2 & a4F7/2 &  1.8099 & 11,446 & 1.12e-03 & 3.28e+04 & 0.20 & 0.20 & 0.20 \\
11 &  2 & a4D5/2 & a6D7/2 &  1.2489 & 12,074 & 3.54e-04 & 2.80e+04 & 0.01 & 0.03 & 0.06 \\
11 &  3 & a4D5/2 & a6D5/2 &  1.2946 & 12,074 & 2.20e-03 & 2.80e+04 & 0.05 & 0.17 & 0.33 \\
11 &  4 & a4D5/2 & a6D3/2 &  1.3281 & 12,074 & 1.30e-03 & 2.80e+04 & 0.03 & 0.10 & 0.19 \\
11 &  6 & a4D5/2 & a4F9/2 &  1.5339 & 12,074 & 2.64e-03 & 2.80e+04 & 0.05 & 0.18 & 0.33 \\
11 &  7 & a4D5/2 & a4F7/2 &  1.6773 & 12,074 & 2.11e-03 & 2.80e+04 & 0.04 & 0.13 & 0.24 \\
11 &  8 & a4D5/2 & a4F5/2 &  1.8005 & 12,074 & 1.55e-03 & 2.80e+04 & 0.03 & 0.09 & 0.17 \\
12 &  4 & a4D3/2 & a6D3/2 &  1.2791 & 12,489 & 2.69e-03 & 2.50e+04 & 0.03 & 0.11 & 0.25 \\
12 &  5 & a4D3/2 & a6D1/2 &  1.2981 & 12,489 & 1.17e-03 & 2.50e+04 & 0.01 & 0.05 & 0.11 \\
12 &  7 & a4D3/2 & a4F7/2 &  1.5999 & 12,489 & 3.53e-03 & 2.50e+04 & 0.03 & 0.11 & 0.26 \\
12 &  8 & a4D3/2 & a4F5/2 &  1.7116 & 12,489 & 9.93e-04 & 2.50e+04 & 0.01 & 0.03 & 0.07 \\
12 &  9 & a4D3/2 & a4F3/2 &  1.7976 & 12,489 & 1.81e-03 & 2.50e+04 & 0.01 & 0.05 & 0.12 \\
13 &  5 & a4D1/2 & a6D1/2 &  1.2707 & 12,729 & 3.60e-03 & 2.80e+04 & 0.02 & 0.07 & 0.16 \\
13 &  8 & a4D1/2 & a4F5/2 &  1.6642 & 12,729 & 4.00e-03 & 2.80e+04 & 0.01 & 0.06 & 0.14 \\
13 &  9 & a4D1/2 & a4F3/2 &  1.7454 & 12,729 & 2.11e-03 & 2.80e+04 & 0.01 & 0.03 & 0.07 \\
14 &  2 & a4P5/2 & a6D7/2 &  0.7640 & 19,387 & 1.17e-02 & 2.13e+05 & 0.05 & 0.10 & 0.32 \\
14 &  6 & a4P5/2 & a4F9/2 &  0.8619 & 19,387 & 2.73e-02 & 2.13e+05 & 0.11 & 0.22 & 0.66 \\
14 &  7 & a4P5/2 & a4F7/2 &  0.9054 & 19,387 & 7.07e-03 & 2.13e+05 & 0.03 & 0.05 & 0.16 \\
15 &  3 & a4P3/2 & a6D5/2 &  0.7689 & 19,673 & 1.22e-02 & 2.12e+05 & 0.03 & 0.05 & 0.21 \\
15 &  7 & a4P3/2 & a4F7/2 &  0.8894 & 19,673 & 1.70e-02 & 2.12e+05 & 0.03 & 0.06 & 0.25 \\
15 &  8 & a4P3/2 & a4F5/2 &  0.9229 & 19,673 & 9.87e-03 & 2.12e+05 & 0.02 & 0.03 & 0.14 \\
16 &  4 & a4P1/2 & a6D3/2 &  0.7667 & 20,006 & 1.13e-02 & 2.28e+05 & 0.00 & 0.02 & 0.09 \\
16 &  8 & a4P1/2 & a4F5/2 &  0.9036 & 20,006 & 1.24e-02 & 2.28e+05 & 0.00 & 0.02 & 0.08 \\
16 &  9 & a4P1/2 & a4F3/2 &  0.9270 & 20,006 & 1.64e-02 & 2.28e+05 & 0.01 & 0.02 & 0.11 \\

\bottomrule
\end{tabular}
\tabnote{
We list all [Fe II] lines resulting from transitions among the levels 
in four ground terms and brighter than 5\%\ of [Fe II] 1.644 $\mu$m line 
in statistical equilibrium at densities $10^3$--$10^5$ cm$^{-3}$ 
and at $T_e=7,000$~K.   
In the table, 
$T_{\rm ex}=$ excitation temperature of the upper level,  
$A_{21}=$Einstein $A$ coefficient,  
$n_{\rm cr}=$critical density at $T_e=7,000$~K, and the  
last three columns = line intensities relative to [Fe II] 1.644 $\mu$m line 
at $n_e=10^3, 10^4$ and $10^5$ cm$^{-3}$ assuming 
statistical equilibrium at $T_e=7,000$~K.
The absolute intensity of [Fe II] 1.644 $\mu$m line 
for $N({\rm H})=10^{20}$~cm$^{-2}$ are 0.00234, 0.0136, and 0.0392 
ergs cm$^{-2}$ s$^{-1}$
when $n_e=10^3, 10^4$ and $10^5$ cm$^{-3}$, respectively,
assuming that Fe abundance is cosmic abundance, 
i.e., $X({\rm Fe/H})=3.47\times 10^{-5}$ in number, and that all Fe is singly ionized.
For the identification of upper and lower levels, 
see the energy diagram in Figure 1. 
}
\end{table*}

The atomic parameters necessary for the calculation of 
\fetwo forbidden lines have been continuously updated with the 
advance of computing power and theoretical modeling.
The Maxwellian-averaged collision strengths for electron-impact excitation have 
been recently calculated by \citet{ramsbottom2007} who included 
the 100 LS terms belonging to the basis configurations 
3d$^6$4s, 3d$^7$, and 3d$^6$4p. They showed that their results differ 
considerably from previous theoretical works \citep[e.g.,][]{zhang1995}.
We find that the ratio of the collision strengths of 
\citet{ramsbottom2007} to those of \cite{zhang1995} 
ranges from 0.35 to 2.3 with a mean of $0.99\pm0.29$ at 5,000 K.  
More recently, \cite{bautista2015} conducted a major study of the [Fe II] atomic
rates, making some new theoretical calculations, comparing the predictions with
those of earlier work, and elucidating the reasons for discrepancies among them.  
Radiative transition probabilities (Einstein $A$ coefficients) in the literature also 
show considerable scatter.
For example, the theoretical $A$-values for the two strongest 
1.257 and 1.644 $\mu$m lines range 4.83--5.27$\times 10^{-3}$ s$^{-1}$ and
4.65--5.07$\times 10^{-3}$ s$^{-1}$, respectively
\citep{nussbaumer1988,quinet1996,deb2010,deb2011,bautista2015}. 
This results in the considerable 
dispersion (1.18--1.36) in the 
expected [Fe II] 1.257/1.644~$\mu$m intensity ratio, which is used to 
derive the extinction \citep{koo2015}. 
On the other hand, \citet{rodrigues2004,smith2006,giannini2015} empirically
derived the ratio between 0.98 and 1.49 from observations of nearby stellar objects.
We will use the effective collision strenghs of \cite{ramsbottom2007} and the $A$ values of 
\cite{deb2011} for the shock code calculation. 
As we will show in \S~2.2, 
the atomic constants of \cite{bautista2015} 
yield line ratios that poorly match the observed ratios for some MIR lines,  
and furthermore they provided effective 
collision strengths only at temperatures above 5,000 K, while, in low-velocity shocks,    
a significant contribution to the MIR [Fe II] lines comes from regions 
at lower temperatures (see \S~3).

Table 1 lists parameters of strong [Fe II] lines 
which are brighter than 5\%\ of the 1.644 $\mu$m line in statistical equilibrium 
at densities $10^3$--$10^5$ cm$^{-3}$ and at $T_e=7,000$~K.   
(For completeness, we have included optical lines resulting from 
the transitions from a$^4$P term.)  
The critical density of level $j$ is defined by
\begin{equation}
n_{\rm cr}\equiv \Sigma A_{ji} (j>i) / \Sigma C_{ij} (j\neq i)  
\end{equation}
where $\Sigma C_{ij}$ is the collisional (de-)excitation 
coefficient averaged over
a Maxwellian-velocity distribution at temperature $T_e$
\citep[e.g.,][]{draine2011}.
The critical densities of NIR [Fe II] lines
are 2.7--3.4$\times 10^4$ cm$^{-3}$ at $T_e=7,000$~K while they are 
lower ($4.8\times 10^2$--$1.7\times 10^4$ cm$^{-3}$) for 
MIR/FIR [Fe II] lines and higher ($2.2\times 10^5$ cm$^{-3}$) for optical lines.

\begin{figure*}
\centering
\includegraphics[width=140mm]{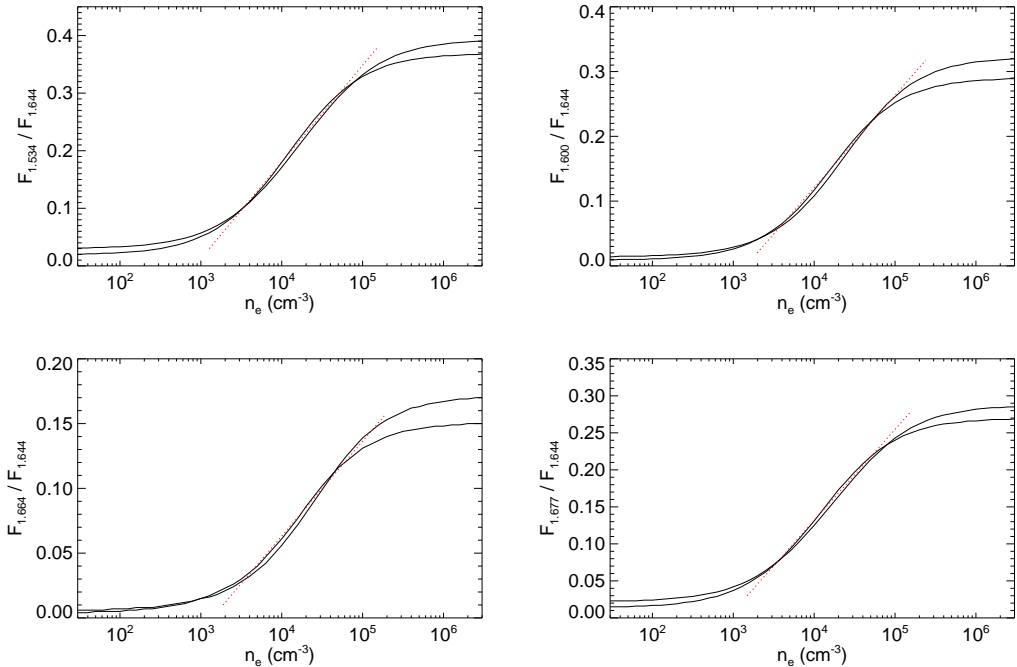}
\caption{
Density sensitive NIR [Fe II] line ratios as a function of electron density for gas 
in statistical equilibrium at temperatures 5,000~K and 10,000~K by thick and thin black lines, respectively. 
The dotted red lines are analytic approximations at 5,000 K (see text).
}
\label{fig2}
\end{figure*}

In order to use the [Fe II] lines to estimate the iron abundance, we need
to know the fraction of Fe in Fe II.  In radiative shock waves, the gas cools
more rapidly from the postshock temperature than the recombination times of
most ions, so that the gas is not in ionization equilibrium.  In addition, 
photoionization of the gas below about 10,000 K dominates the ionization
state in shocks faster than about 100 \kms.  The most important parameters
are the Fe II collisional ionization rate, which \cite{arnaud1992} determined
from the laboratory measurements of \cite{montague1984}, the Fe II photoionization 
cross section, which we take from \cite{reilman1979}
and the Fe III dielectronic and radiative recombination rates, for which we
use \cite{arnaud1992}, and the Fe III+H charge transfer rate \citep{neufeld1987}.
The photoionization cross section we use is about 20\% smaller than that of
\cite{verner1996}, which is within the uncertainties.
The dielectronic recombination rates differ significantly from those given by
\cite{nahar1997}, but the charge transfer rate dominates in the regions where
it matters.

\subsection{Basic Applications}

\subsubsection{NIR [Fe II] lines}

One of the most useful applications of the NIR [Fe II] lines
is to measure extinction. 
There are lines originating from the same upper levels, 
the ratios of which are functions of only 
the Einstein $A$ coefficients (and their wavelengths).
Their observed 
ratios provide an accurate measure of extinction to the emitting region.
The two strong \fetwo lines at 1.257~ and 1.644 $\mu$m are such lines.
If we adopt $A_{1.257}/A_{1.644}=1.04$ \citep{deb2011}, 
the intrinsic ratio of their line intensities 
$j_{1.257}/j_{1.644}=1.36$, so that  
the extinction (in mag) difference 
at 1.257 and 1.644 $\mu$m ($\Delta A_{JH}$) can be derived from 
\begin {equation} 
\Delta A_{JH}=1.086\log\left( F(1.257)/F(1.644) \over [F(1.257)/F(1.644)]_{\rm int} \right) 
\end{equation}
where $[F(1.257)/F(1.644)]_{\rm int}(=1.36)$ is the intrinsic line flux 
ratio. According to \cite{bautista2015}, 
the uncertainty in the intrinsic ratio is $\sim 20$\%, which yields 
an uncertainty of 0.09 mag in $\Delta A_{JH}$. 
This corresponds to $A_V\approx 0.9$ mag for the Galactic 
extinction curve with $R_V=3.1$.  
There are a couple of observable NIR lines 
sharing the same upper state, e.g., 1.321 $\mu$m 
and 1.372 $\mu$m lines (see Figure 1 and Table 1). 
There are also 
[Fe II] lines at 0.90--0.95 $\mu$m sharing their upper levels 
but they are weak (see Figure 1 and Table 1).


\begin{figure*}[!b]
\stepcounter{figure}
\centering
\includegraphics[width=140mm]{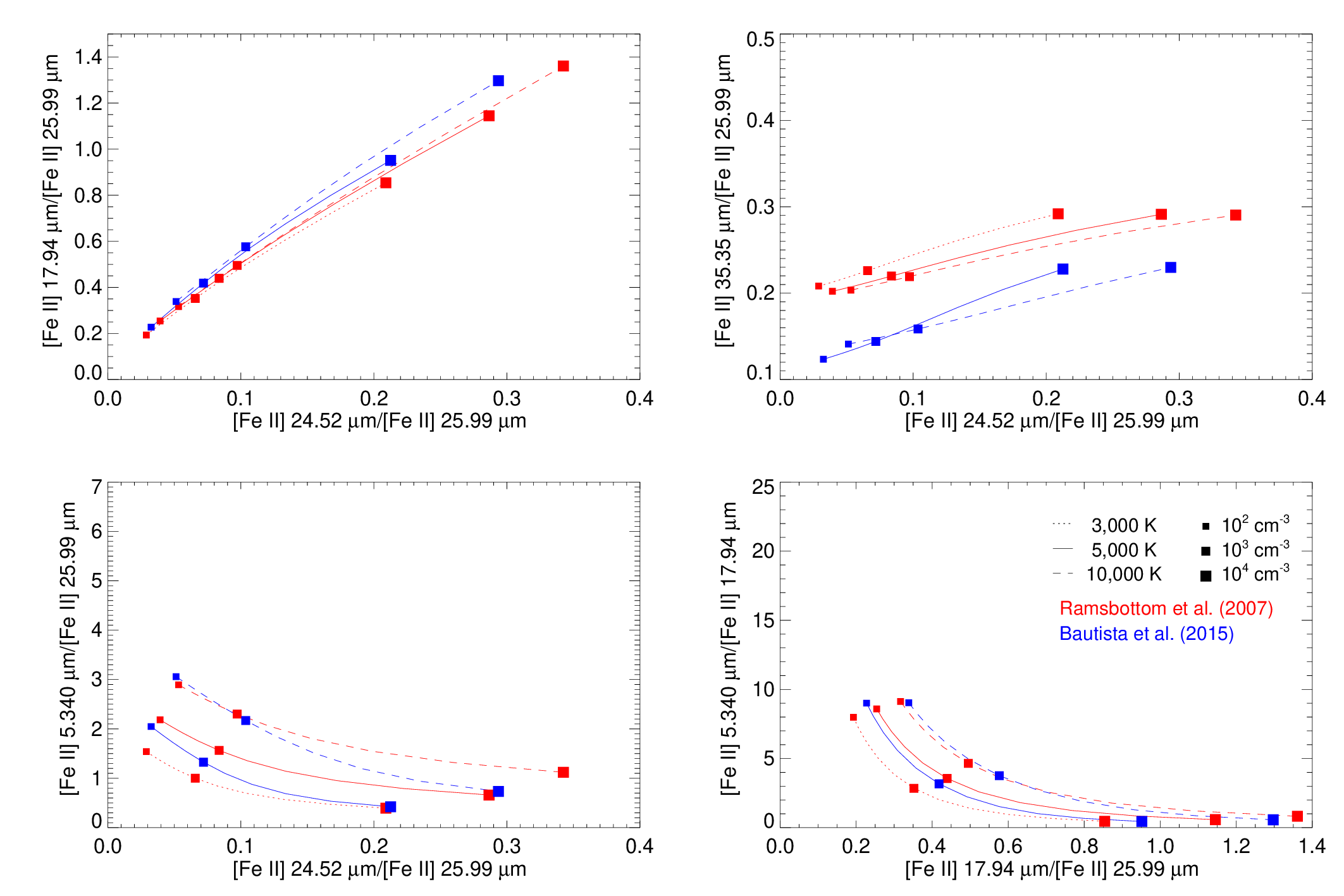}
\caption{
MIR [Fe II] line intensity ratios for thermal gas in 
statistical equilbirum at $T=3,000$, 5,000, and 10,000 K.  
Along each line, squares mark electron densities of  
$10^2$, $10^3$, and $10^4$~cm$^{-3}$. 
The red lines represent the 
results obtained by using the effective collision sterengths of 
\cite{ramsbottom2007} and the $A$ values of \cite{deb2011}, while 
the blue lines are the results obtained by using those of \cite{bautista2015}.}
\label{fig4}
\end{figure*}

\begin{figure}[t]
\addtocounter{figure}{-2}
\centering
\includegraphics[width=80mm]{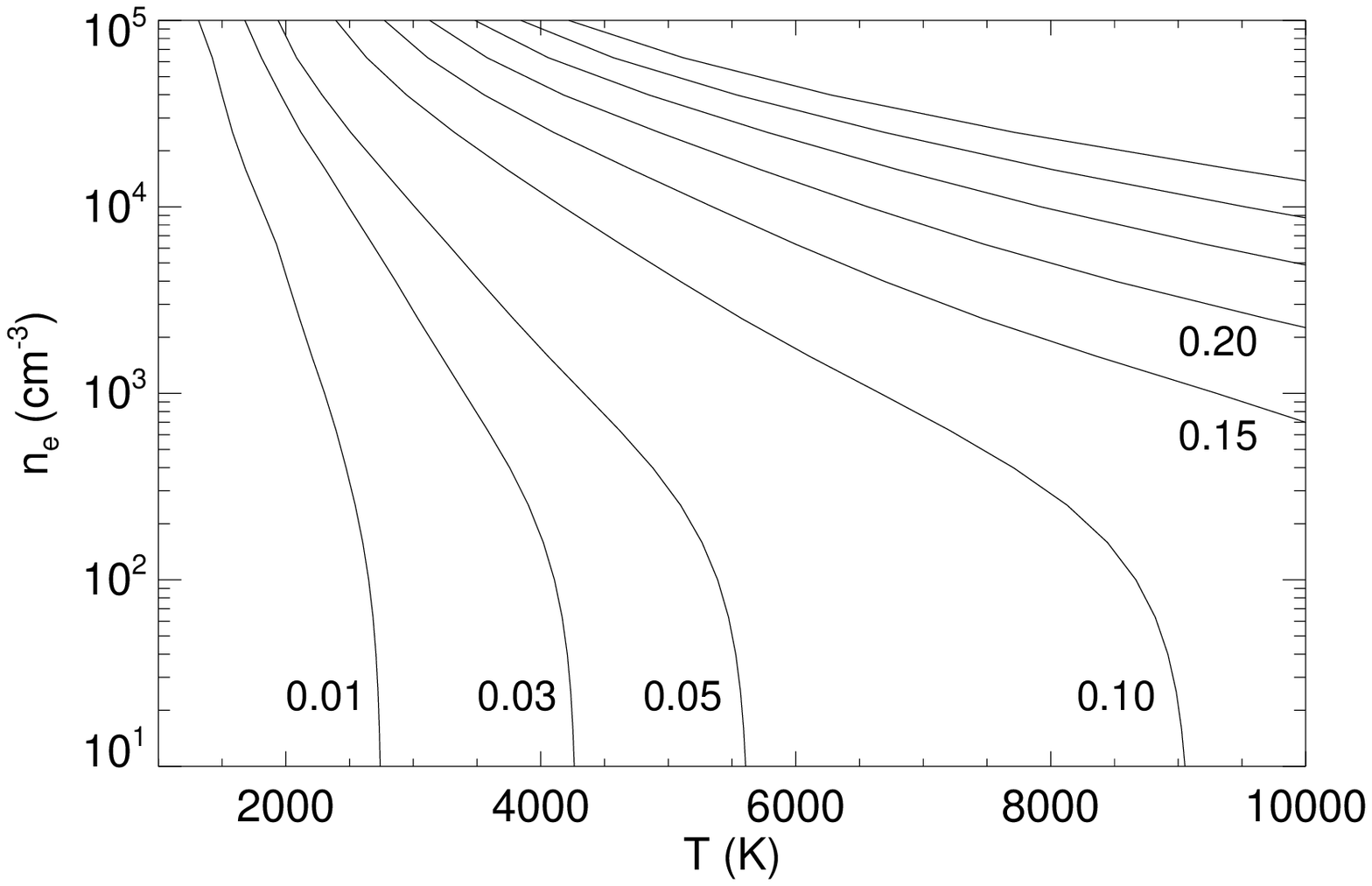}
\caption{
[Fe II] 0.8619 $\mu$m/[Fe II] 1.257 $\mu$m intensity ratios 
of thermal gas in statistical equilibrium in $(T, n_e)$ plane.   
The contours of constant intensity ratios are plotted. Above 0.20, 
the contours are equally spaced in steps of 0.05.  
}
\label{fig3}
\addtocounter{figure}{1}
\end{figure}

Another useful application is density diagnosis. 
There are several lines of comparable excitation energies, 
the ratios of which are mainly a function of electron density ($n_e$)
of the emitting region, depending only weakly on temperature ($T_e$).
Figure 2 shows four representative line ratios 
$r_{1.534/1.644}\equiv F(1.534)/F(1.644)$,
 $r_{1.600/1.644}\equiv F(1.600)/F(1.644)$,
 $r_{1.664/1.644}\equiv F(1.664)/F(1.644)$,
and $r_{1.677/1.644}\equiv F(1.677)/F(1.644)$ 
for gas at temperature 5,000~K and 10,000~K. 
At low densities ($n_e\ll n_{\rm cr}$), their ratios are equal to the
ratio of collisional excitation rates, while 
at high densities ($n_e\gg n_{\rm cr}$), their ratios 
are given by the ratio of spontaneous deexcitation rates.
These line ratios can be used as a density diagnostic 
for $n_e=10^{3-5}$ cm$^{-3}$. 
Convenient linear approximations that can be used for 
$n_e=3\times 10^3$ to $8\times 10^4$ cm$^{-3}$ are   
\begin{equation} 
n_e = 10^{2.93+5.95 r_{1.534/1.644}},~~~~~   0.10\le r_{1.534/1.644} \le 0.32 
\end{equation}
\begin{equation} 
n_e = 10^{3.16+7.01 r_{1.600/1.644}},~~~~~   0.06\le r_{1.600/1.644} \le 0.24
\end{equation}
\begin{equation} 
n_e = 10^{3.14+13.7 r_{1.664/1.644}},~~~~~   0.03\le r_{1.664/1.644} \le 0.13 \\
\end{equation}
\begin{equation} 
n_e = 10^{2.93+8.14 r_{1.677/1.644}},~~~~~   0.07\le r_{1.677/1.644} \le 0.23 \\
\end{equation}
The maximum error of the above fits in electron density is less than 15\%.

If the electron density is known from the above, then the ratio of [Fe II] lines with 
considerably different excitation energies can be used 
for temperature diagnosis, although this requires observations of [Fe II] lines in different bands.
Figure 3 is a diagnostic diagram using the ratio of 0.8619 $\mu$m and 1.257 $\mu$m lines, which are the  
two strongest lines from a$^4$P and a$^4$D terms to a$^6$D term, respectively.
This may be compared to similar plots in previous studies, e.g., Figure 8 of \cite{graham1990}. 
In shocked gas, these lines originate from a region 
at $\sim 7,000$~K (see \S~3.2), so the temperature diagnosis using [Fe II] lines is not 
particularly useful for interstellar shocks.

\subsubsection{MIR [Fe II] lines}

MIR [Fe II] lines are not useful for extinction measurent. There are 5.340 $\mu$m and 6.722 $\mu$m lines sharing 
the upper level (a$^4$F$_{9/2}$), but the 6.722 $\mu$m line is weak, i.e., 
their intrinsic intensity ratio $F(6.722)/F(5.340)=0.071$ (Table 1). 
Also, the effect on MIR line intensities will be small unless the 
extinction is very large.

Intensity ratios of MIR [Fe II] lines may be used for density diagnosis, but they are 
rather temperature sensitive. In Figure 4, we plot the contours of constant 
temperatures at $n_e=10^2$ to $10^4$~cm$^{-3}$ in the plane of the 
intensity ratios of strong MIR lines. 
The dependence of ratios on $n_e$ and $T$ can be understood by noting that  
the 25.99 $\mu$m and 35.35 $\mu$m lines have comparable 
\tex\ (550--960 K) and \ncr\ (2.3--$3.4\times 10^3$~cm$^{-3}$), while 
the 17.94 $\mu$m and 24.52 $\mu$m lines have comparable 
\tex\ (3,500--4,100 K) and \ncr\ (1.1--$1.5\times 10^4$~cm$^{-3}$)
that are higher than those of the former lines (Table 1).
Therefore, the intensity ratio of the latter to the former lines 
will increase with density as well as with temperature. 
On the other hand, the 5.340 $\mu$m line has a much lower critical density 
\ncr ($=4.4\times 10^2$~cm$^{-3}$) and \tex\ of 2,700 K,
so that the intensity ratio of 5.340 $\mu$m to the other MIR lines will decrease with density.
In addition to these, Figure 4 shows that the ratios 
$F(17.94)/F(25.99)$ and $F(24.52)/F(25.99)$
are proportional each other, nearly independent of temperature and density (upper left frame), 
implying $F(24.52)/F(17.94)\sim 0.4$. 

In Figure 4, the red lines represent the 
results obtained by using the effective collision sterengths of 
\cite{ramsbottom2007} and the $A$ values of \cite{deb2011}, while 
the blue lines are the results obtained by using those of 
\cite{bautista2015}\footnote{\cite{bautista2015} provided 
the effective collision strengths and the branching ratios for 
radiative transitions in machine-readable forms in Tables 12 and 10, 
respectively. The numerical values in these tables, however, are not their 
recommended values but the 7-config model values in Table 8 
and the TFDAc values in Table 4, respectivly.}.
We note that the two sets of atomic constants 
yield $F(24.52)/F(25.99)$ ratios that are comparable at small $n_e$ but  
considerably, e.g., by a factor of 1.4 at 5,000 K, different at $n_e=10^4$~cm$^{-3}$. 
They also yield $F(35.35)/F(25.99)$ ratios that differ  
by a factor of $\sim 1.5$ across most of the density range. 
These differences between the two results should be mostly due to the differences 
in collision strengths because the radiative transition rates for strong lines are 
almost the same except for a$^4$F$_{9/2}$ level where the 5.340 $\mu$m line originates. 
For the 5.340 $\mu$m line, the $A$-value of \cite{bautista2015} 
is smaller than that of \cite{deb2011} by a factor of 2. 
The collision strengths of \cite{bautista2015} are generally 
smaller than those of \cite{ramsbottom2007} by a factor of $\simlt 3$ (see their Table 8).
As we will see in \S~4, 
the result of \cite{ramsbottom2007} is in better agreement with 
the observations, but it also poorly matches the observed ratios involving 
the 5.340 $\mu$m line.


\section{Radiative Atomic Shocks and IR [Fe II] Emission}

\subsection{Shock Code and Model Parameters}

The shock code that we use is the one 
developed by \citet{raymond1979} and improved by \citet{cox1985},  
to which we have added the [Fe II] model described above.  It
solves the equations for a steady flow of the shocked gas as
it cools, computing the time-dependent ionization state including
photoionization. The line emissivities $j_\nu$ (ergs cm$^{-3}$ s$^{-1}$ sr$^{-1}$) 
of a fluid element are then calculated 
following its trajectory from the shock front until it cools down to 
1,000 K or lower.
These emissivities are integrated along the line of sight to obtain 
the line flux normal to the shock front, i.e.,
\begin{equation}
F_\nu =2\pi \int j_\nu dx.
\end{equation}

For shocks slower than about 110~\kms, the temperature jump is 
sensitive to the ionization fraction of preshock gas. 
If a substantial fraction of H atom is neutral, 
the effective temperature jump is reduced because a 
considerable fraction of shock energy is used in exciting and ionizing H atoms, 
which makes the H line intensities stronger. 
For the preshock ionization fraction of H, we adopt the result of 
\cite{shull1979}. 
For He and heavier elements, we first calculate the shock structures assuming that 
they are in photoionization equilibrium with the interstellar radiation field,  
and then use the resulting shock radiation to 
calculate the ionization fraction of the preshock gas.
The temperature structure at the shock front 
does not significantly affect the intensities of [Fe II] lines but does affect 
the ratios of the [Fe II] lines to the H lines. 

We consider shock speeds $v_s=20$ to 200 \kms\ and 
preshock densities of H nuclei from $n_0$=10 to 1000 cm$^{-3}$.
The magnetic field strength $B_0$ limits the 
maximum density in the postshock layer and therefore the line 
intensity ratios.  The [Fe II] lines are mainly 
emitted in the layer with moderate compression, and their 
ratios are not directly sensitive to magnetic field strength,
though stronger fields imply lower densities and higher
ionization states in the photoionized region where the [Fe II]
lines form.  According to \cite{heiles2005} and \cite{crutcher2010}, 
the magnetic field strength in the diffuse 
ISM ($n\le 300$~cm$^{-3}$) does not scale with density, 
and the median total magnetic field strength is 6 $\mu$G. 
We adopt 5~$\mu$G as the tangential magnetic field strength in our models. 
We adopt the solar abundances by \cite{asplund2009}.

The models do not include emission from the photoionization precursor,
which becomes important for shocks faster than about 150 \kms \citep{dopita1996},
although slower shocks in principle could produce an extensive precursor where
Ly$\alpha$ ionizes Fe I to Fe II. 
However, the iron is strongly depeleted
ahead of the shock, and the precursor will contribute little to the [Fe II]
emission.  In the case of fast shocks in the LMC, where the precursor is
not spatially resolved from the postshock flow, there may be significant
hydrogen Pa$\beta$ (1.282 $\mu$m) from the precursor.

\subsection{Shock Structure and [Fe II] Emission}

In radiative atomic shocks, an extended region of partially ionized gas 
develops behind 
the shocks, and this is where the [Fe II] emission originates. 
Below we describe the structure of radiative shocks and explore 
the emission characteristics of [Fe II] lines for 
two shocks of $v_s= 150$ and 30 \kms\   
propagating into a medium of 
$n_0=10^2$~cm$^{-3}$ and $n_0=10^3$~cm$^{-3}$, respectively. 
The preshock gas is fully ionized in the former shock, while it 
is neutral in the latter shock.

Figure 5 (top frame) shows the temperature profile 
in the postshock cooling layer of the 150~\kms\ shock. 
The corresponding profiles of H nuclei and electron densities 
are shown in the bottom frame. 
At $N_{\rm H}\sim 7\times 10^{17}$~cm$^{-2}$ 
the cooling becomes important and 
temperature abruptly drops to 8,000 K. 
Then the temperature remains roughly constant over an extended region to 
$N_{\rm H}\sim 1\times 10^{19}$~cm$^{-2}$. 
This temperature plateau region is essentially an 
H II region where the heating is provided by UV radiation 
generated from the hot gas just behind the shock front.
In contrast to HII regions around OB stars 
where H atoms are essentially fully ionized, however, 
the H ionization fraction is relatively low.
Note that the gas density increases slowly in the plateau region 
while the electron density decreases logarithmically as 
the gas recombines.
For the detailed description of the structure of radiative shocks, 
the readers may refer \citet[][and references therein]{shull1979}.

\begin{figure}[t!]
\centering
\includegraphics[width=80mm]{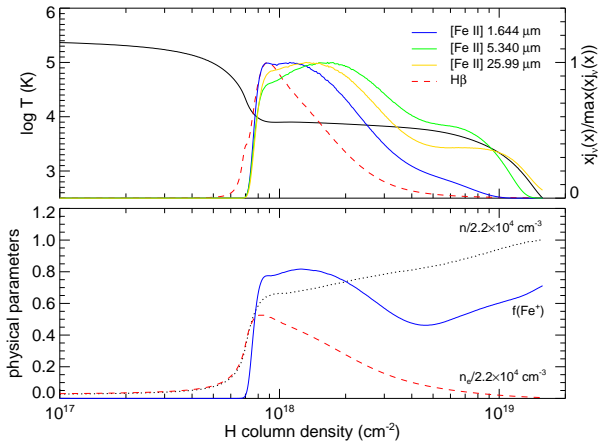}
\vspace{-5mm}
\caption{Structure of 150~\kms\ shock 
propagating into ambient medium of $n_0=10^2$~cm$^{-3}$ and $B_0=5$~$\mu$G.
The abscissa is H-nuclei column density swept up by shock. 
In the top frame, black solid line shows the 
temperature profile in logarithmic scale. Also shown are the 
normalized emissivities of [Fe II] and H$\beta$ lines in linear scale.
Note that we plot $xj_\nu(x)$ because the abscissa is in logarithmic scale.
In the bottom frame, we plot the profiles of H nuclei density ($n$), 
electron density ($n_e$) and fraction of Fe in Fe$^+$ (Fe$^+$/Fe). 
}
\label{fig5}
\end{figure}

In the bottom frame of Figure 5, we also plot the profile of 
Fe$^+$ fraction and, in the top frame, the 
emissivities of the [Fe II] 1.644, 5.340, 25.99 $\mu$m and H$\beta$ lines. 
Note that the Fe$^+$ fraction is high and remains flat (0.4--0.8) far 
downstream, even reaching the region where $T\sim 1,000$~K. 
The high Fe$^+$ fraction at $N_{\rm H}\simlt 5\times 10^{18}$~cm$^{-2}$,
i.e., in the region before the kink in 
the Fe$^+$ fraction profile, is due to collisions with electrons and to charge
transfer with H atoms. 
But in the region beyond that, 
the gas ionization fraction is low and  
collisions with electrons become less important. 
Here, since the ionization potential of iron atoms is 7.9 eV,
FUV photons from the hot shocked gas
can penetrate far downstream to
maintain the ionization state of Fe$^+$
where H atoms are primarily neutral.
But as can be seen in the top frame, 
this extended Fe$^+$ region produced by FUV radiation 
does not contribute much to the [Fe II] forbidden line 
emission because of low electron density. 
Most of the NIR [Fe II] emission originates from the 
temperature plateau region where the ionization fraction is 
not too low. The emissivity-weighted 
temperature and electron density of 
[Fe II] 1.644 $\mu$m (1.534 $\mu$m) line are 7,100 K 
(7,500 K) and 6,200 cm$^{-3}$ (7,200 cm$^{-3}$), respectively. 
For comparison, the electron density that would have been derived from 
the ratio $F(1.534)/F(1.644) (=0.13)$ is 
5,400 cm$^{-3}$, somewhat lower than these.  The
[Fe II] 5.340 $\mu$m and 25.99 $\mu$m lines, whose critical densities are lower, are  
emitted from a more extended region than the NIR [Fe II] line emitting region.
The emissivity-weighted mean temperature and electron density 
of these lines are $\sim 5,800$ K and $\sim 4,000$~cm$^{-3}$.
In Figure 5 we also show the emissivity 
of H$\beta$ line. Note that the [Fe II] lines are emitted 
in a more extended region than the H$\beta$ line, which is one of the reasons that 
the ratio of [Fe II] to H$\beta$ lines from shocked gas is much 
higher than that in photoionized H II regions.

\begin{figure}
\centering
\includegraphics[width=80mm]{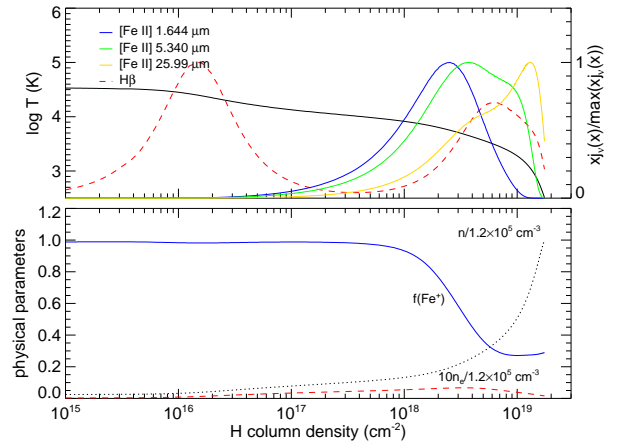}
\vspace{-5mm}
\caption{
Same as Figure 5 but for 30~\kms\ shock 
propagating into ambient medium of $n_0=10^3$~cm$^{-3}$ and $B_0=5$~$\mu$G.
}
\label{fig6}
\end{figure}

Figure 6 is the same plot for a 30~\kms\ shock. The temperature profile is not very different from 
that of the 150~\kms\ shock, except that we do not see 
an abrupt temperature drop and the resulting temperature plateau. 
This is because the incoming hydrogen is neutral, and  
a significant fraction of thermal energy is lost to its excitation. 
The strong H$\beta$ emission at $\simlt 10^{17}$~cm$^{-2}$ is from the collisionally 
excited neutral hydrogen. This emission is indeed stronger than that 
from the recombining H in the far downstream.   
Behind the shock front, Fe is mostly in Fe$^+$. (It was Fe$^{+2}$ for the 150~\kms\ shock.)  
[Fe II] lines are emitted where the fraction of Fe$^+$ is high and also 
the electron density is high. The [Fe II] 5.340 $\mu$m and 25.99 $\mu$m line  
emitting regions are shifted farther downstream compared to the  150~\kms\ shock case. 
(We have extended the calculation until the temperature drops to 300 K because significant 
25.99 $\mu$m emission originates from the gas at $T_e\le 1,000$~K.)   
The emissivity-weighted temperatures for the 1.644, 5.340, and 25.99 $\mu$m lines are 
7000, 5700, and 4400 K, respectively. The corresponding electron densities are 
700, 680, and 630 cm$^{-3}$, respectively. The 
ratio $F(1.534)/F(1.644)$ is 0.045, which would yield 
$n_e=700$ cm$^{-3}$. 

\subsection{IR [Fe II] Line fluxes and Comparison with Other Shock Models}

\begin{figure}[t!]
\includegraphics[width=80mm]{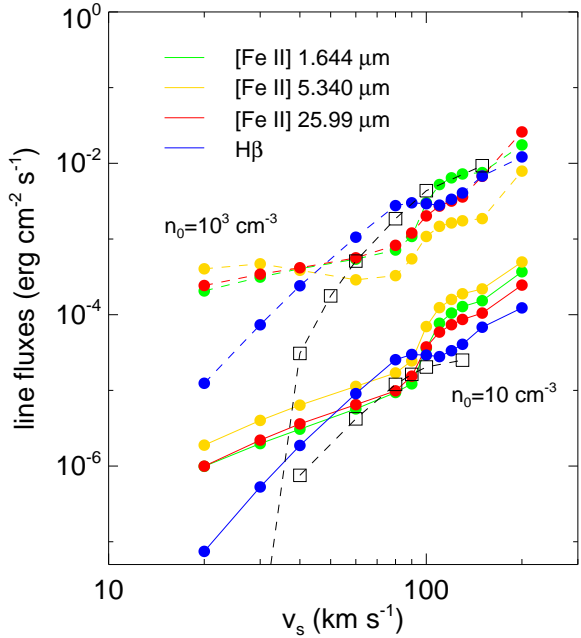}
\caption{
[Fe II] line fluxes normal to the shock front as a function of shock speeds 
for preshock densities 10 and $10^3$ cm$^{-3}$. 
H$\beta$ fluxes are also plotted for comparison. The filled circles 
along a constant preshock density line represent $v_s= 
20, 30, 40, 60, 80, 90, 100, 110, 120, 130, 150,$ and 
200~\kms. The empty squares are H$\beta$ fluxes 
of \cite{hollenbach1989} and \cite{shull1979} 
for $n_0=10$ and $10^3$ cm$^{-3}$, respectively.
}
\label{fig7}
\end{figure}

Figure 7 shows the fluxes of \feonesix, 5.340 \mum, 25.99 $\mu$m, and \hbeta\ 
lines as a function of shock speed when $n_0=10$ and $10^3$~cm$^{-3}$. 
If a constant fraction of the total shock kinetic energy flux 
$\rho_0 v_s^3/2$ is converted into a line radiation, the line flux will be
proportional to $v_s^3$. Indeed, \hbeta\ fluxes at $v_s\ge 110$~\kms\ are well described by 
\begin{equation}
F_{\rm H\beta} 
\simeq 3.5\times 10^{-4} \left(f_{\rm H\beta} \over 0.00575\right) n_{0,2} 
v_{s,7}^3 ~~{\rm erg~cm}^{-2}~{\rm s}^{-1} {\rm sr}^{-1}.
\end{equation}
where $f_{\rm H\beta}(\approx 0.6$\%) is the fraction of incoming shock energy converted to \hbeta\ line flux. 
The plateau in \hbeta\ fluxes between about 70 and 110 $\rm km~s^{-1}$
results from the collisional excitation contribution to \hbeta\ when
the preshock gas is partially neutral.
At low shock speeds, a significant fraction of the shock energy 
is used to ionize the incoming neutral H, but the temperatures are low enough
that more of the energy goes into Ly$\alpha$ and less into \hbeta, so that 
the line flux drops faster \citep{raymond1979} 
\citep[see also Fig. 11 of ][]{hollenbach1989}.
The incoming hydrogen atoms are fully ionized at  
shock speeds $\ge 110$--120 \kms\ independent of preshock density.
In Figure 7, we compare our \hbeta\ results to those of \citet{hollenbach1989}
and \cite{shull1979} for $n_0=10^3$ and 10 cm$^{-3}$, respectively.
The consistency among the different shock models is generally good.
The line fluxes of \citet{hollenbach1989} are systematically lower than ours   
at low shock speeds because significant shock energy goes to 
H$_2$ dissociation and excitation, while we assume that the preshock
gas is atomic. 

Figure 7 shows that the [Fe II] lines are 
stronger than the \hbeta\ line over most of the shock speed range 
when $n_0=10$~cm$^{-3}$ and at 
low shock speeds when $n_0=10^3$~cm$^{-3}$. 
The kink at $\sim$100~\kms\ is  
probably because the UV radiation at the shock front 
becomes stronger, increasing the electron density 
in the [Fe II] emitting region. 

We have also compared our results to those predicted by 
MAPPINGS III \citep{allen2008}. 
For a shock with $v_s=150$~\kms, $n_0=100$~cm$^{-3}$, and $B_0=5$~$\mu$G 
for example, our code gives $F(1.644)/F({\rm H} \beta$) of 2.1  
and $F(1.257)/F({\rm Pa} \beta$) of 16  
which are a factor of 5 greater than 
those (0.40 and 3.1) of the MAPPINGS III.
We note that the profiles of physical parameters for the same 
shock parameters are quite comparable except that  
the cooling starts later 
and the cooling region extends further in the MAPPINGS III code.
Apparently the extent of the temperature plateau region where the
[Fe II] lines originate is considerably ($\sim 1/2$) narrower in the 
MAPPINGS III case. We find that 
about half of the difference appears to originate from 
different choices of atomic rates.
The other half may result from the treatment
of the radiative transfer in the resonance lines. 

\begin{figure}[t!]
\includegraphics[width=80mm]{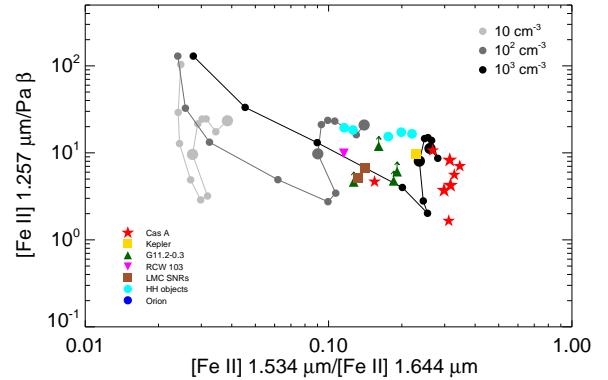}
\caption{
[Fe II] 1.257 $\mu$m/Pa$\beta$ vs. 
[Fe II] 1.534 $\mu$m/[Fe II] 1.644 $\mu$m diagram. 
The lines of constant preshock 
densities are shown for $n_0=10, 10^2,$ and $10^3$~cm$^{-3}$. 
Along each line, circles mark shock speeds of  
$20, 30, 40, 60, 80, 90, 100, 110, 120, 130, 150,$ and 
200~\kms\ with 100 and 200~\kms\ marked by bigger circles. 
The abundance is solar.   
}
\label{fig8}
\end{figure}

\begin{table*}[t!]
\caption{Observed NIR [Fe II] line ratios in SNRs and HH objects}
\centering
\begin{tabular}{lccccccc}
\toprule
{Name} & \multicolumn{3}{c}{\underline {\hfil ~~~~~~~~~~~~~~Observed~~~~~~~~~~~~~~ \hfil}} & 
{$A_V$} & \multicolumn{2}{c}{\underline{\hfil ~~~~~Dereddened~~~~~ \hfil}} \\
{} & {$\displaystyle {1.257 \over 1.644}$} & {$\displaystyle {1.534 \over 1.644}$} & 
{$\displaystyle {1.257\over {\rm Pa}\beta}$} & {(mag)} & 
{$\displaystyle {1.534 \over 1.644}$} & {$\displaystyle {1.257\over {\rm Pa}\beta}$} & Ref  \\
\midrule
\underline{Galactic SNRs}\hfil &&&&&&&\\
       Kepler &   1.21 (0.00) &   0.22 (0.00) &    ...        &    1.3 &   0.23 &   9.70 &  1  \\
 G11.2-0.3 C1 &   0.30 (0.02) &   0.14 (0.01) &  $> 5.32$     &   16.5 &   0.19 &  $> 6.08$ & 2   \\
 G11.2-0.3 C2 &   0.26 (0.03) &   0.13 (0.03) &  $> 4.11$     &   18.2 &   0.18 &  $> 4.76$ & 2   \\
 G11.2-0.3 C3 &   0.22 (0.02) &   0.09 (0.01) &  $> 3.95$     &   19.7 &   0.13 &  $> 4.63$ & 2   \\
    G11.2-0.3 &   0.31 (0.01) &   0.12 (0.00) &  $> 10.47$    &   16.1 &   0.16 & $> 11.92$ & 3   \\
        Cas A &   0.85 (0.00) &   0.29 (0.00) &    ...        &    5.1 &   0.32 &   4.24 & 1   \\
        Cas A &   1.67 (0.00) &   0.30 (0.00) &    ...        &    0.0 &   0.30 &   3.71 & 1   \\
        Cas A &   0.73 (0.00) &   0.27 (0.00) &    ...        &    6.8 &   0.31 &   8.25 & 1   \\
        Cas A &   0.80 (0.01) &   0.14 (0.02) &   4.45 (0.22) &    5.8 &   0.15 &   4.67 & 4   \\
        Cas A &   0.64 (0.00) &   0.26 (0.02) &   1.55 (0.02) &    8.2 &   0.31 &   1.65 & 4   \\
        Cas A &   0.64 (0.01) &   0.29 (0.08) &   6.59 (0.65) &    8.3 &   0.35 &   7.05 & 4   \\
        Cas A &   0.63 (0.01) &   0.23 (0.02) &   9.95 (0.73) &    8.4 &   0.27 &  10.66 & 4   \\
        Cas A &   0.63 (0.02) &   0.28 (0.02) &   5.22 (0.37) &    8.4 &   0.33 &   5.59 & 4   \\
      RCW 103 &   0.88 (0.04) &   0.10 (0.01) &   9.50 (2.66) &    4.8 &   0.12 &   9.88 & 5   \\
\\       
\underline{LMC SNRs}\hfil &&&&&&&\\
          N49 &   1.33 (0.00) &   0.14 (0.03) &   6.65 (0.00) &    0.2 &   0.14 &   6.66 & 6   \\
         N63A &   1.26 (0.00) &   0.13 (0.03) &   5.04 (0.00) &    0.8 &   0.13 &   5.07 & 6   \\
\\         
\underline{HH objets}\hfil &&&&&&&\\
       HH111F &   0.55 (0.01) &   0.14 (0.01) &  14.20 (4.26) &    9.9 &   0.18 &  15.38 & 7   \\
       HH111H &   0.57 (0.01) &   0.10 (0.01) &  18.00 (6.01) &    9.5 &   0.12 &  19.43 & 7   \\
       HH240A &   1.06 (0.01) &   0.21 (0.01) &  16.19 (1.01) &    2.8 &   0.22 &  16.55 & 7   \\
       HH241A &   0.81 (0.03) &   0.11 (0.02) &  17.50 (5.85) &    5.6 &   0.13 &  18.32 & 7   \\
        HH120 &   0.99 (0.02) &   0.19 (0.01) &  16.80 (4.20) &    3.5 &   0.20 &  17.28 & 7   \\

        
\bottomrule
\end{tabular}
\tabnote{
Observed and dereddened line ratios. `...' means not detected. 
The errors in parenthesis are usually $1\sigma$ statistical errors. `(0.00)' means no errors are given. 
Observed line ratios are dereddened assuming the intrinsic ratio 
[Fe II] 1.257 $\mu$m/[Fe II] 1.644 $\mu$m = 1.36 and by using the 
Galactic extinction curve with $R_V=3.1$. For the sources with 
the observed [Fe II] 1.257 $\mu$m/[Fe II] 1.644 $\mu$m ratio $> 1.36$,
we adopted $A_V=0$.
The derived visual extinctions ($A_V$) are listed in the fifth column. \\
} 
\tabnote{Comments on indivdual sources:\\
Kepler: Pa$\beta$ was not identified by \cite{gerardy2001} but Pa$\gamma$ was. 
        We have derived  
        [Fe II] 1.257 $\mu$m/Pa$\beta$ assuming Pa$\beta$/Pa$\gamma$=1.83 
        which is the ratio of Case B at $T_e=7,000$~K. \\
G11.2$-$0.3: We have taken $3\sigma$ as an upper limit for Pa$\beta$. \\ 
Cas A: Same as the Kepler SNR. \\
N63A and N49: [Fe II] 1.534 $\mu$m line had not been observed by 
\cite{oliva1990}. But they observed [Fe II] 1.600 $\mu$m line and obtained 
[Fe II] 1.600 $\mu$m/[Fe II] 1.644 $\mu$m 
$=0.09\pm 0.02$ and $0.08\pm 0.02$, which correponds to 
[Fe II] 1.534 $\mu$m/[Fe II] 1.644 $\mu$m $=0.14$ and 0.13 at 
$T_e=7,000$~K.\\
}
\tabnote{References for the observed line ratios:
(1) \cite{gerardy2001}; (2) \cite{lee2013}; (3) \cite{koo2007}; (4) \cite{koo2013};
(5) \cite{oliva1990}; (6) \cite{oliva1989,oliva2001}; 
(7) \cite{nisini2002}
}
\end{table*}

\begin{table*}[t!]
\caption{Observed MIR [Fe II] line ratios for SNRs and HH objects}
\centering
\begin{tabular}{lccccccc}
\toprule
{Name} & {$\displaystyle {35.35 \over 25.99}$} & {$\displaystyle {24.52 \over 25.99}$} & 
{$\displaystyle {17.94\over 25.99}$} & {$\displaystyle {5.340 \over 25.99}$} 
& {$\displaystyle {5.340\over 17.94}$} & 
IRS mode & Ref  \\
\midrule
\underline{Galactic SNRs}\hfil &&&&&&&\\
  Cygnus Loop &   0.27 (0.00) &   0.10 (0.00) &   0.27 (0.00) &  NA &  NA &  SH, LH & 1 \\
       Kes 69 &  NA &   0.09 (0.01) &   0.25 (0.01) &   3.45 (0.22) &  13.88 (0.99) &  SL, LL & 2 \\
        3C396 &  NA &   0.14 (0.03) &   0.46 (0.01) &   3.08 (0.06) &   6.66 (0.12) &  SL, LL & 2 \\
       Kes 17 &  NA &   0.06 (0.00) &   0.31 (0.02) &   3.66 (0.17) &  11.68 (0.81) &  SL, LL & 2 \\
   G346.6-0.2 &  NA &   ... &   ... &   ... &   ... &  SL, LL & 2 \\
   G348.5-0.0 &  NA &   0.09 (0.00) &   0.38 (0.00) &   5.02 (0.10) &  13.15 (0.26) &  SL, LL & 2 \\
   G349.7+0.2 &  NA &   0.20 (0.00) &   0.51 (0.00) &   1.68 (0.02) &   3.28 (0.03) &  SL, LL & 2 \\
          W44 &   0.24 (0.00) &   0.03 (0.01) &   0.21 (0.00) &   4.78 (0.00) &  22.77 (0.00) &  SL, SH, LH & 3 \\
          W28 &   0.26 (0.00) &   0.06 (0.00) &   0.19 (0.00) &   3.38 (0.00) &  17.36 (0.00) &  SL, SH, LH & 3 \\
        3C391 &   0.29 (0.00) &   0.08 (0.00) &   0.26 (0.00) &   2.42 (0.00) &   9.24 (0.00) &  SL, SH, LH & 3 \\
        IC443 &   0.91 (0.00) &  $<$0.11      &   0.22 (0.00) &   2.80 (0.44) &  12.64 (2.00) &  SL, SH, LH & 3 \\
\\
\underline{LMC SNRs}\hfil &&&&&&&\\   
      N63A SW &   ...         &   0.28 (0.00) &   0.49 (0.01) &   1.32 (0.16) &   2.67 (0.32) &  All  & 4 \\
      N63A NE &   0.30 (0.00) &   0.20 (0.00) &   0.70 (0.00) &   1.13 (0.04) &   1.62 (0.06) &  All  & 4 \\
      N63A SE &   0.29 (0.00) &   0.20 (0.00) &   0.53 (0.00) &   1.33 (0.10) &   2.52 (0.19) &  All  & 4 \\
\\
\underline{HH objects}\hfil &&&&&&&\\
       HH54FS &   0.28 (0.05) &   0.07 (0.01) &   0.26 (0.05) &   2.58 (0.24) &   9.77 (2.20) &  SL, SH, LH & 5 \\
        HH54C &   0.27 (0.07) &   0.09 (0.01) &   0.33 (0.05) &   4.08 (0.47) &  12.55 (2.33) &  SL, SH, LH & 5 \\
       HH54EK &   0.41 (0.06) &   0.14 (0.02) &    $<$0.36    &   1.88 (0.72) &    $>5.20$    &  SL, SH, LH & 5 \\
          HH7 &   0.34 (0.04) &    $<$0.07    &    $<$0.24    &    ...        &    ...        &  SL, SH, LH & 5 \\
    SMM1 blue &   0.30 (0.04) &   0.35 (0.02) &   0.26 (0.01) &     NA        &    NA         &  SH, LH &  6 \\
     SMM1 red &   0.28 (0.05) &   ... &   0.15 (0.01) &  NA &  NA &  SH, LH &  6 \\

\bottomrule
\end{tabular}
\tabnote{
The numbers are surface brightness ratios. `NA' means not observed, while `...' means not detected. 
The errors in parenthesis are usually $1\sigma$ statistical errors. `(0.00)' means no errors are given. 
The systematic errors in the Spitzer IRS fluxes havae been estimated to be $\le 25$\% \citep{neufeld2006}, so
the undertainties in the flux ratios would be $\le 35$~\%.
Note that the lines obtained by different 
Spitzer IRS modules are from different areas, i.e.,  
SH (9.9 - 19.6) and LH (18.7 - 37.2) modules in high resolution mode and 
SL (5.2-14.5) and LL (13.9-39.9) modules in low resolution mode.\\
}
\tabnote{Comments on indivudal sources:\\
Hewitt sources: \cite{hewitt2009} did not separate the [Fe II] 
25.99 $\mu$m line from the nearby [O IV] line, and they did 
not give any numbers for the [Fe II] 35.35 $\mu$m line.  
However, the 35.35 $\mu$m line is clearly seen on the wing of the [Si II]
line in Kes79, 3C396, G48.5-0.0 and G349.7+0.2 at about 1/3 the strength
of the 25.99 $\mu$m line. This plus the absence of any [Ne V] emission indicates
that the 25.99 $\mu$m feature is dominated by [Fe II], though the presence of
[Ne III] emission suggests that some [O IV] emission may be present. \\
}
\tabnote{References: (1) \cite{sankrit2014}; 
(2) \cite{hewitt2009}; (3) \cite{neufeld2007}; (4) \cite{caulet2012}; (5) \cite{neufeld2006}; (6) \cite{dionatos2014} 
}
\end{table*}

\section{Shock Grids and Comparison to Observations}
\subsection{NIR [Fe II] Lines}

Figure 8 is a diagram of $F(1.257)/F({\rm Pa}\beta)$ vs. 
$F(1.534)/F(1.644)$ where the grids of constant preshock densities are plotted.
The $F(1.257)/F({\rm Pa}\beta)$ ratio has a kink between 80 and 130 \kms\ 
because the Pa$\beta$ flux remains flat in that velocity range (see Figure 7).   
 
In Figure 8, we also plot the line ratios observed toward SNRs
and Herbig-Haro (HH) objects (see Table 2). 
The SNR data points are believed to be associated with shocks propagating into 
the ambient medium, as we exclude observations of ejecta knots, although there 
could be some contamination from SN material.  In order for comparison, we 
applied the extinction correction using the observed $F(1.257)/F(1.644)$ 
ratios. We also assumed some basic principles, e.g., the 
$F({\rm Pa}\beta)/F({\rm Pa}\gamma)$ ratio is given by that of Case B, if necessary (see the note in Table 2).     
The observed 
$F(1.257)/F({\rm Pa}\beta)$ ratios range from 2 to 20. For comparison, 
it is 0.02--0.03 in Orion bar \citep{walmsley2000}. 

Figure 8 shows that the observed line ratios toward SNRs and HH objects are 
well explained by shocks propagating at $v_s=80$--200~\kms\ into a medium of 
$n_0=10^2$--$10^3$~cm$^{-3}$ with solar abundances. 
(The two LMC points need to be shifted upwards in the diagram in order to be  compared 
to the models because the Fe abundance in the LMC is about 1/2 of the solar abundance \cite{russell1992}.)  
According to Figure 8, the Cas A SNR shocks are experiencing the highest 
preshock densities of about $1,000$~cm$^{-3}$, while 
the shocks in HH objects appear to have the fastest ($\simgt 150$~\kms) 
shock speeds. The fact that the observations
match models with solar abundances suggests nearly complete liberation of 
Fe from dust grains in these shocks.

\subsection{MIR [Fe II] Lines}

There are {\it Spitzer} MIR spectroscopic observations of various 
interstellar shocks. Table 3 lists the MIR [Fe II] line intensity ratios 
of some sources available in literature. 
They include SNRs and HH objects. 
The 25.99 $\mu$m line is the bright line between the lowest
two levels and is available in all the
spectra, so we have normalized the line intensities by the 25.99 $\mu$m line.
The {\it Spitzer} IRS spectrometer, however, is composed of two, 
short and long wavelength, modules sampling different parts of the sky.
Therefore, the ratios of the lines obtained from different modules 
could be far from what we would expect. In some cases, the emitting region
was mapped by rastering the slits \citep{caulet2012, dionatos2014}, 
and in others it is possible to scale the
fluxes in the different apertures by using lines in overlapping wavelength ranges of 
LL and SL modules \citep{hewitt2009} or by using lines in the different wavelength ranges that 
have known intensity ratios \citep{sankrit2014}. We exclude ratios
involving lines observed in different modules where neither method
of normalization is possible.

Figure 9 is a diagram of $F(17.94)/F(25.99)$ vs. 
$F(24.52)/F(25.99)$.  
As we noted in \S~3, the 17.94 $\mu$m line has nearly the same
\tex\ and \ncr\ as the 24.52 $\mu$m line, 
so the models lie along a straight line.
(See the upper left frame in Figure 4. 
The plots in this section use the same line ratios 
as those in Figure 4.)  
The observed intensities are generally below 
the models, especially  
the protostellar jet SMM1 and the LMC SNR N63A SW.
There is a considerable degeneracy in deriving shock speeds 
and/or preshock densities from these line ratios. 
For example,
the observed $F(17.94)/F(25.99)$ ratio ($=0.27$) of the Cygnus Loop SNR would imply either
$(n_0,v_s)\approx (10~{\rm cm}^{-3}, 120~{\rm km~s}^{-1})$ 
or $(10^2~{\rm cm}^{-3}, 45~{\rm km~s}^{-1})$ 
or $(10^3~{\rm cm}^{-3}, 25~{\rm km~s}^{-1})$. According to a detailed shock modeling of 
the IR plus optical/UV spectra of the Cygnus Loop by \cite{sankrit2014}, 
a 150~\kms\ shock propagating into a medium of $n_0\sim 5$~cm$^{-3}$ 
can reasonably explain the observed spectra, so the first set of parameters is 
close to the right solution.

\begin{figure}[t!]
\includegraphics[width=80mm]{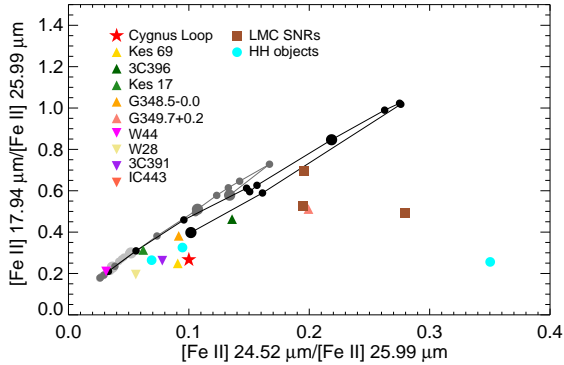}
\caption{
[Fe II] 17.94 $\mu$m/[Fe II] 25.99 $\mu$m vs. 
[Fe II] 24.52 $\mu$m/[Fe II] 25.99 $\mu$m diagram. 
The shock grids are the same as in Figure 8.
}
\label{fig9}
\end{figure}
\begin{figure}
\includegraphics[width=80mm]{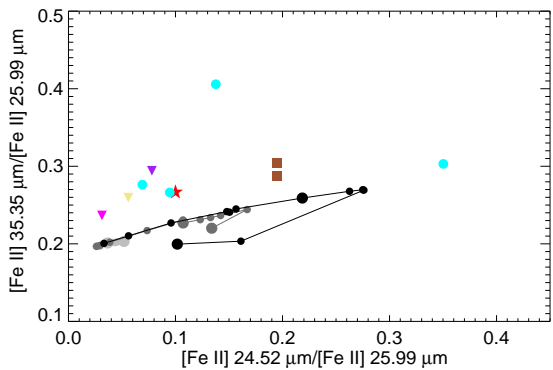}
\caption{
Same as Figure 9 but for 
[Fe II] 35.35 $\mu$m/[Fe II] 25.99 $\mu$m vs. 
[Fe II] 24.52 $\mu$m/[Fe II] 25.99 $\mu$m.  
}
\label{fig10}
\end{figure}

Figure 10 is a diagram of $F(35.35)/F(25.99)$ vs. 
$F(24.52)/F(25.99)$.  Here the observed ratios lie above
the predictions by $\simlt 30$~\% over most of the 
$F(24.52)/F(25.99)$ ratio range, 
though one HH object lies much higher. 
Apparently, the result of the 
statistical equilibrium calculation appears to be in a 
better agreement with the observed $F(35.35)/F(25.99)$ ratios 
at high $F(24.52)/F(25.99)$ ratios   
(upper right frame in Figure 4). 
The poor agreement with the statistical equilbrium results  
at low $F(24.52)/F(25.99)$ ratios suggests that 
there are some fundamental issues in 
the $F(35.35)/F(25.99)$ ratios at low densities. 
The shock models using the coefficients of \cite{bautista2015} yield 
an even larger offset as we can infer from Figure 4.
 
Figure 11 shows $F(5.340)/F(25.99)$ vs. $F(24.52)/F(25.99)$. 
Here again, the observed $F(5.340)/$ $F(25.99)$ ratios lie above
the models, especially for low $F(24.52)/F(25.99)$ ratios.
Figure 4 suggests that the presence of Fe$^+$ at higher temperatures 
would improve the agreement. But again 
the agreement is poor (by a factor of $\sim 2$) at 
low $F(24.52)/F(25.99)$ ratios, which seems to indicate that  
there are some fundamental issues in 
the $F(5.340)/F(25.99)$ ratios at low densities.  
Figure 12 presents $F(5.340)/F(17.94)$ vs. $F(17.94)/F(25.99)$,
and it shows that although the 5.340 $\mu$m and 17.94 $\mu$m lines arise
from neighboring levels the agreement is not better.

\begin{figure}
\includegraphics[width=80mm]{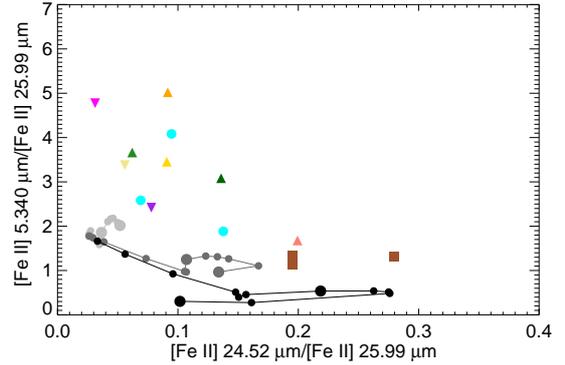}
\caption{
Same as Figure 9 but for 
[Fe II] 5.340 $\mu$m/[Fe II] 25.99 $\mu$m vs. 
[Fe II] 24.52 $\mu$m/[Fe II] 25.99 $\mu$m. 
}
\label{fig11}
\end{figure}

\begin{figure}
\includegraphics[width=80mm]{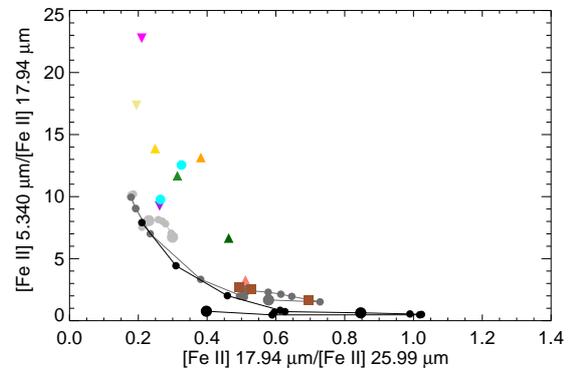}
\caption{
Same as Figure 9 but for 
[Fe II] 5.340 $\mu$m/[Fe II] 17.94 $\mu$m vs. 
[Fe II] 17.94 $\mu$m/[Fe II] 25.99 $\mu$m. 
}
\label{fig12}
\end{figure}

\section{Discussion}

The results of the comparison of [Fe II] line intensity ratios predicted from 
the shock models to those observed in astronomical shocks 
may be summarized as follows.  
First, in the $F(1.257)/F({\rm Pa}\beta)$ vs. $F(1.534)/F(1.644)$ plane, 
the observed ratios fall on the model grids with 
reasonable range of shock parameters (Figure 8). 
So just based on this plot, the shock models appear to 
yield reasonably accurate relative fluxes of NIR [Fe II] lines. 
Second, among the MIR line intensity ratios, 
the $F(24.52)/F(25.99)$ and $F(17.94)/F(25.99)$ ratios 
predicted from the shock models both cover the range of the 
observed ratios but do not yield 
consistent results for most data points
(Figure 9). 
Third, $F(35.35)/F(25.99)$ ratios predicted from the shock models 
are smaller than the observed ratios by $\simlt 30$\% 
for most of the density range (Figure 10).
Fourth, $F(5.340)/F(25.99)$ and $(F5.340)/F(17.94)$ ratios  
predicted from the shock models 
are significantly (by up to a factor of $\simlt 5$) smaller 
than the observed ratios at all
densities (Figures 11 and 12).
In the following, we discuss the
possible contributions of errors in the atomic physics, the shock
wave models, and the observations to the problems.

\subsection{Atomic physics}

The first source of errors that we can think is 
an uncertainty in atomic parameters.
\cite{bautista2015} present several new calculations of the [Fe II]
atomic rates and compare them with earlier calculations. They recommend
the average of their computations, which is largely determined by four
calculations using the \cite{quinet1996} target configurations 
for radiative transition rates. 
According to \cite{bautista2015}, the uncertainties in the 
radiative transition rates of the lowest 16 levels are 
generally better than 10~\%. 
The a$^4$F$_{9/2}$ level where the 5.340 $\mu$m line originates 
has an exceptionally large uncertainty of 30\%. 
The recommended values for collision strengths among the lowest states have 
an rms uncertainty of 10--50~\%, and they are roughly
half as large those given their DARC model or the R-MATRIX calculation
of \cite{ramsbottom2007}.
If the Einstein $A$ coefficient of the 5.340 $\mu$m line 
were greater, the $F(5.340)/F(25.99)$ and $F(5.340)/F(17.94)$ ratios 
would have been in a better agreement with the observations. 
The largest offset between the predicted and the observed 
ratios in $F(5.340)/F(25.99)$ and $(F5.340)/F(17.94)$ are found 
at low density ends, which suggests an error 
in collision strengths rather than $A$ coefficients.   

One potential problem is that our model includes only 16 energy levels,
and cascades from higher levels that are not included might influence the
line intensities.  We have used the CHIANTI package \citep{landi2013}, which
includes many higher energy levels, to estimate this contribution. (CHIANTI
version 7 uses Fe II collision strengths from \cite{zhang1995}, which are
similar to those of \cite{ramsbottom2007}.) At
$10^4$ K, cascades contribute around 30\% to the population of the upper
levels of the lines we observe, but much of that comes from the levels
that we include, and the contribution from higher levels would be smaller
at the temperatures of 4,000 to 7,000 K where the bulk of the emission
arises. We conclude that cascades from higher energy levels not included
in the model probably do not affect the ratios considered here at more
than the 20\% level.

\subsection{Shock models}

The shock models are idealized in many ways.  They assume a single shock
speed and steady flow from the shock until the gas reaches 1,000 K.
What matters for the ratio-ratio plots is the distribution of Fe$^+$
ionization fraction over the density and temperature of the emitting
region. Since the models span a reasonably broad range of parameters,
they probably cover real parameters fairly well, even in cases such as
the LMC SNRs, where the observed region undoubtably includes a range
of shock speeds. If the real shocks are incomplete, meaning that the
gas does not have time to cool all the way to 1,000 K, the cool part
of the Fe II emitting region would be absent, and that could affect the
line ratios, but one would also expect places where the hotter part of
the emitting region predicted in steady flow would be absent.

The models ignore emission from the shock photoionization precursor, but
since the Fe is probably highly depleted ahead of the shock, that should be
a good approximation. The radiation transfer in the model ignores resonant
scattering, which is likely to be important for Ly$\alpha$ and the He I and
He II resonance lines.  Comparison with the MAPPINGS III model suggests that
this may affect on the shock structure at a level as large as that caused
by the atomic rate uncertainties, but it is not clear whether it would
systematically change the ratio-ratio diagrams.

An interesting possible process not included in any model is that Fe atoms
liberated from grains in the hot postshock gas will emit some [Fe II]
photons before they are ionized to Fe III.  The process has been observed
in UV emission from C IV \citep{raymond2013}, but the emission is very faint.  The
relatively low excitation rates of forbidden lines and the high ionization
rate Fe II make it unlikely that this process contributes significantly
to the observed fluxes.

\subsection{Observational uncertainties}

Extinction affects the NIR [Fe II] and Pa$\beta$ lines, typically by
10\% to 30\% (e.g., see Table 2), and it is unlikely that 
errors in the extinction correction is a major contributor to the uncertainty. 
The {\it Spitzer} observations use different modules for 
the 5.340 $\mu$m and 17.94 $\mu$m lines than
for the 25.99 $\mu$m and 35.35 $\mu$m lines, which could present a substantial uncertainty.
We have included only observations where the region was mapped so that the
flux from same area could be extracted, or where scaling of the channels
using a known line ratio was possible.  Moreover, the ratio-ratio plots
show trends, such as good agreement at high densities and poor agreement at
low densities, that are unlikely to arise from inconsistent treatment of
the different channels. This argument would not apply to the outlying
point at the top of Figure 10 or perhaps the HH object point at the lower
right of Figure 9, however.

\section{Conclusions}

We find that the biggest obstacle to the interpretation of [Fe II] line
intensities in the NIR and MIR ranges is still the uncertainty in the
atomic rates for this complex ion. Neither of the most advanced calculations
of collision strengths gives a good match to the ratio-ratio diagrams of MIR lines,
and the predicted ranges for some MIR line ratios are significantly offset 
from the observed ratios. While uncertainties in the shock modeling and the observations certainly exist, 
they seem less likely to resolve the problems. 

The NIR line ratios can provide extinction to the source 
with an uncertainty of $A_V\approx 0.9$~mag  
and density of emitting region with an uncertainty of about 20\%. 
The shock parameters derived from NIR line ratios might be reliable.  
The MIR line ratios, however, are sensitive both to density and temperature, 
although the density could be constrained to some range from 
ratios such as $F(17.94)/F(25.99)$ and $F(24.52)/F(25.99)$
The shock parameters may be obtained from MIR line ratios, 
but one should be cautious. 

The most important application of the [Fe II] IR lines for interstellar
shocks is probably the derivation of the iron abundance, often giving an
indication of the fraction of refractory grains destroyed in the shock.
Models predict that the mass liberated from silicate grains increases
from near zero for a 50 $\rm km~s^{-1}$ shock to around 40\% for a
150 $\rm km~s^{-1}$ radiative shock \citep{slavin2015}.  The ratio of
the 1.257 $\mu$m line to Pa$\beta$ is consistent with this prediction if most of the
shocks are faster than about 80 $\rm km~s^{-1}$. An indication of shock
speeds would be the presence or absence of the [S III] line at 9532 \AA.
Of the objects in Table 2, it is present in HH 240 and 241 and absent in
HH 111 and 120. The gas phase abundance of iron can be estimated
from the ratios of the NIR [Fe II] lines to [P II] 1.189 $\mu$m line or 
MIR [Fe II] lines to the [Ne II] 12.82 $\mu$m line 
\citep[e.g.,][]{giannini2008, koo2013, sankrit2014}, since phosphorus and neon
are not depleted onto grains.
\cite{sankrit2014} estimated an iron abundance
around 1/2 solar in a 150 $\rm km~s^{-1}$ shock in the Cygnus Loop, but
there was considerable spread in the ratios of observed 
to predicted intensities for the [Fe II] lines.

\acknowledgments

This research was supported by Basic Science Research Program through the National Research
Foundation of Korea(NRF) funded by the Ministry of Science, ICT and future Planning
(2014R1A2A2A01002811). H.-J. Kim was supported by
NRF(National Research Foundation of Korea) Grant funded by the Korean Government 
(NRF-2012-Fostering Core Leaders of the Future Basic Science Program).

\end{document}